\documentclass[article,superscriptaddress,showkeys,showpacs,longbibliography]{revtex4-1}
\usepackage[utf8]{inputenc}
\usepackage{graphicx}
\usepackage{amssymb, amsmath}
\usepackage{color,soul}
\usepackage{textcomp}
\usepackage{float}

\begin{document}
 
\title{Pearl-Necklace-Like Local Ordering Drives Polypeptide Collapse}
\author{Suman Majumder}
\affiliation{Institut f\"{u}r Theoretische Physik, Universit\"{a}t Leipzig, Postfach 100 920, 04009 Leipzig, Germany}

\author{Ulrich H. E. Hansmann}
\affiliation{Department of Chemistry and Biochemistry, University of Oklahoma, Norman, Oklahoma 73019, USA}

\author{Wolfhard Janke}
\affiliation{Institut f\"{u}r Theoretische Physik, Universit\"{a}t Leipzig, Postfach 100 920, 04009 Leipzig, Germany}
\date{\today}
 
\begin{abstract}
\textbf{Collapse of the polypeptide backbone is an integral part of  protein folding. Using polyglycine 
as a probe, we explore the nonequilibrium pathways of protein collapse in water. 
We find that the collapse depends on the competition between hydration effects and intra-peptide interactions. 
Once intra-peptide van der Waal interactions dominate, the chain collapses along a nonequilibrium 
pathway characterized by formation of pearl-necklace-like local clusters as intermediates 
that eventually coagulate into a single globule. By describing this coarsening through the 
contact probability as a function of distance along the chain, we   extract a time-dependent 
length scale that grows in linear fashion. The collapse dynamics is characterized by a dynamical 
critical exponent $z=0.5$ that is much smaller than the values of $z=1-2$ reported for  non-biological
polymers. This difference in the exponents is explained  by the instantaneous formation of intra-chain hydrogen bonds
and local ordering that may be correlated with the observed fast folding times 
in proteins.}
\end{abstract}
 
\maketitle
Changing the solvent condition from good to poor renders an extended polymer to undergo a collapse transition by forming
a compact globule \cite{stockmayer1960,nishio1979}. Both experiments \cite{Pollack2001,Sadqi2003} 
and simulations \cite{camacho1993,reddy2017} indicate that a protein also experiences such a collapse transition
while folding into its native state. However, the nonequilibrium dynamics of the collapse of 
proteins is only poorly understood and an active research topic \cite{asthagiri2017}. Most previous studies consider
only the  hydrophobicity of apolar side chains of amino acids in  a protein as driving force for collapse  \cite{kauzmann1959,Dill1990}.
In the present paper we focus instead on the contributions  by intra-peptide interactions,
 present even for residues with no hydrophobic  or only  weakly hydrophobic side chains \cite{Bolen2008,Tran2008,Holthauzen2010,teufel2011}
where the collapse-driving forces are not necessarily proportional to the exposed surface. Our test system is polyglycine,
and has been chosen to connect our work with recent studies of
homopolymer collapse dynamics \cite{MajumderEPL,Majumder2016PRE,majumder2017SM,christiansen2017JCP} that found nonequilibrium
scaling laws as known for generic coarsening phenomena \cite{Bray_article}. Our hope is to establish such scaling laws also for 
the collapse of proteins.  As a first stride towards this goal, here, we explore the kinetics of collapse of polyglycine.
\par
Collapse of homopolymers was first described by de Gennes' seminal 
``sausage'' model  \cite{deGennes1985}, but today  the phenomenological 
``pearl-necklace'' picture by Halperin and Goldbart \cite{Halperin2000} is more commonly used, both for flexible \cite{byrne2000,Abrams2002,yeomans2005,reddy2006,guo2011,MajumderEPL,majumder2017SM,christiansen2017JCP}
and semiflexible polymer models \cite{Montesi2004,lappala2013}. In this picture the  
collapse begins with nucleation of small local clusters (of monomers) leading to formation of an 
 interconnected chain of (pseudo-)stable clusters, i.e., the  
``pearl-necklace'' intermediate. These clusters grow by eating up the un-clustered monomers from the chain
 and subsequently coalesce,  
leading eventually to a single cluster. Finally, monomers within this final  cluster rearrange to form a compact globule.
\par
Of central interest  in this context is the scaling of the collapse time $\tau_c$ with the degree of polymerization $N$ (the number of monomers). 
While scaling of the form $\tau_c \sim N^z$, where $z$ is the dynamic exponent, has been firmly established, there is no consensus on the 
value of $z$. Molecular dynamics (MD) simulations provide much smaller values ($z \approx 1$) than Monte Carlo (MC) simulations ($z \approx 2$).
This difference is often explained with the 
presence of hydrodynamics in the MD simulations, but a value  $z \approx 1 $ has been reported recently also for MC simulations  \cite{majumder2017SM}.
The ``pearl-necklace'' stage or the cluster-growth kinetics can be understood by monitoring the 
time ($t$) dependence of the mean cluster size $C_s(t)$, the relevant length scale. By drawing analogy with coarsening ferromagnets,  it has been shown 
that scaling of the form $C_s(t) \sim t^{\alpha_c}$ with growth exponent $\alpha_c \approx 1$ holds for flexible homopolymers \cite{MajumderEPL,majumder2017SM}. 
\par
Protein collapse is much less understood. While it has been shown by modeling a protein as semiflexible heteropolymer  
that  the equilibrium scaling of the radius of gyration $R_g$ with $N$ is random-coil-like in a good solvent and  
globule-like in a  poor solvent  \cite{wilkins1999,uversky2002}, there have been few attempts to explore  nonequilibrium collapse 
pathways  \cite{cooke2003,pham2010}, and the corresponding scaling laws are  not known.
In order to probe the existence of such nonequilibrium scaling laws in protein collapse, we have simulated polyglycine chains
$\textrm{(Gly)}_N$ of various numbers $N$ of residues. This choice allows us    to probe in a systematic way
the collapse of polypeptide chain, considering only homopolymers built from the simplest amino acid, glycine. Our results 
show that in water there is a {\em tug of war} between collapse-disfavoring hydration effects and collapse-favoring intra-peptide interactions. 
For longer chains ($N \ge 15$) the intra-peptide interactions win over the hydration effect leading to a collapse, making  
water in practice a poor solvent. We use these  longer polyglycine chains to shed 
light on the collapse kinetics, with an emphasis on the presence of nonequilibrium scaling laws.
Our results from all-atom MD simulations in the NVT ensemble using a hydrodynamics preserving thermostat (see the Method section for details), 
suggest a collapse mechanism that relies on fast local ordering by formation of pearl-necklace structures which eventually merge into
a single globule. This process is characterized by a dynamic critical exponent $z=0.5$ much smaller than the exponents $z=1-2$ observed for
non-biological polymers, and we speculate that this quicker local ordering and collapse enables the fast folding times seen in proteins.

\par
\begin{figure}
\centering
\includegraphics[width=0.6\textwidth]{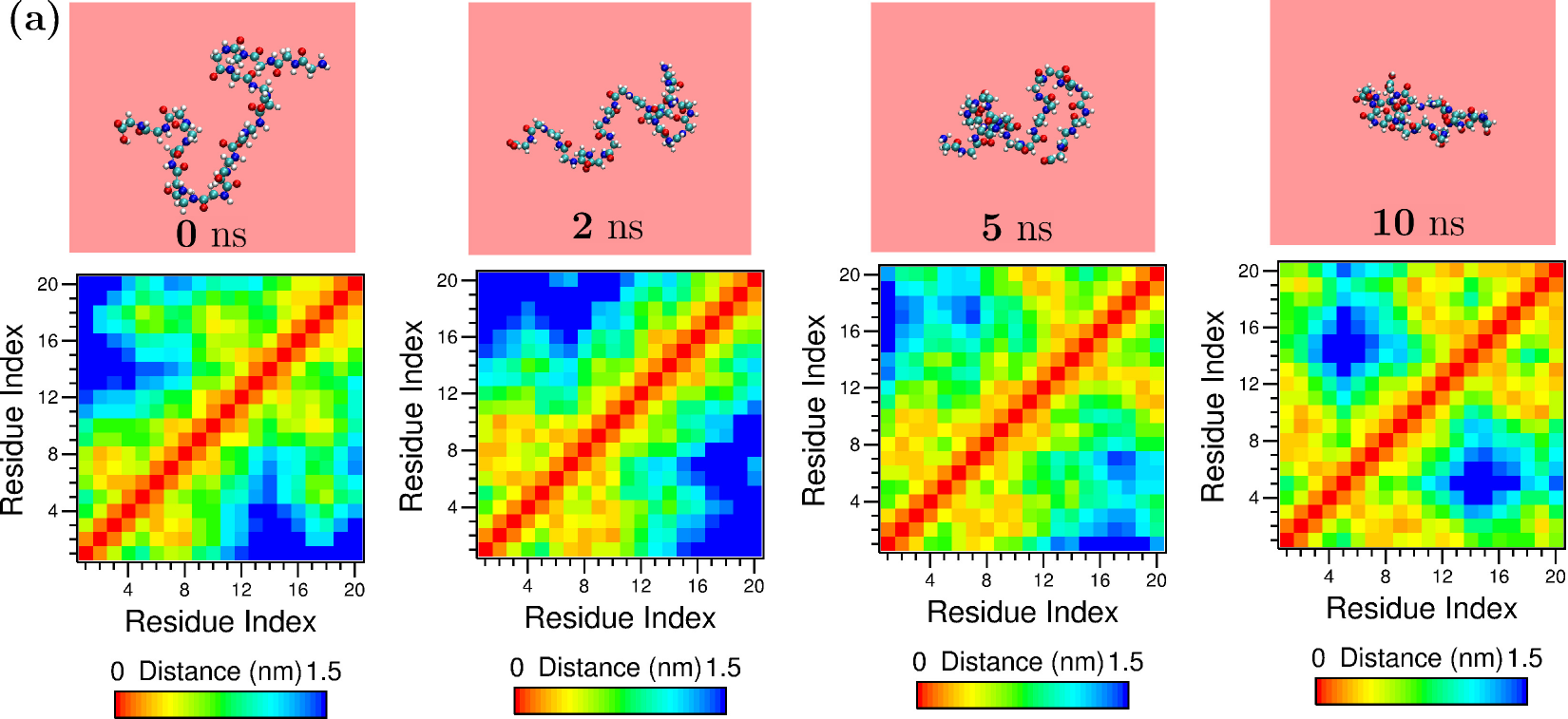}
\includegraphics[width=0.29\textwidth]{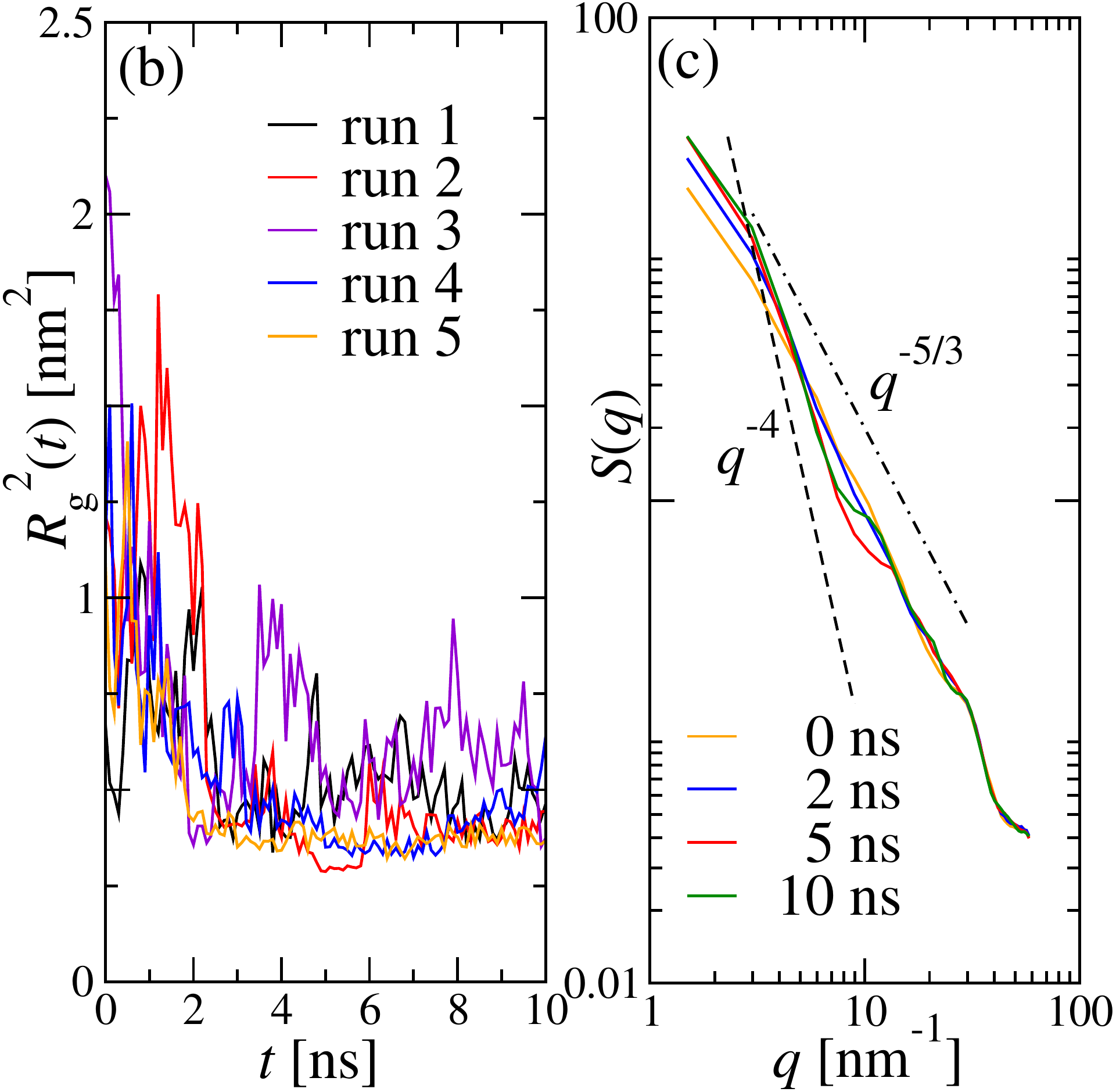}
\caption{\textbf{Time evolution of a short polypeptide.} a) The upper panel shows snapshots from the time evolution for collapse of $\textrm{(Gly)}_{20}$ chain in water at $T_q=290$ K, 
starting from an extended state at $t=0$ ns. The lower panel shows the corresponding residue contact maps where two residues along the chain are in 
contact if the distance between them is less than $1.5$ nm. (b) Time dependence of the squared radius of gyration $R_g^2(t)$ from five different runs. 
(c) Illustration of the structural evolution of the chain during the collapse shown via structure factor $S(q)$ as function of the wave vector $q$ at the 
times presented in (a). The dashed lines with power-law decay exponent $5/3$ and $4$ correspond
to the expected behavior for an extended chain and crumpled globule, respectively. }
\label{snap_GL20}
\end{figure} 

\par
We begin our analysis with a rather short chain, i.e., $\textrm{(Gly)}_{20}$. The time evolution snapshots during the collapse in water at 
a temperature $T_q=290K$, well below the corresponding collapse transition temperature, are shown
 in Fig.\ \ref{snap_GL20}(a). In a protein, collapse leads eventually to folding characterized by formation of distinct 
native contacts among the residues. We show for this reason in the lower panel the residue contact maps where we define  
two residues as being  in contact if they are within a distance of $r_c=1.5$ nm. The red stripe along the diagonals depicts the self-contacts. 
The size of the extended  $\textrm{(Gly)}_{20}$ chain is $\approx 2.0$ nm, thus almost all the mutual distances between the residues fall under $r_c$. 
This makes it difficult to capture segregation or formation of any local structures on length scales comparable to $r_c$. 
Only late in the trajectories do we find a signature for loop formation,  which is also apparent in the snapshot at $t=10$ ns. 
Emergence of such loop is due
 a competition between the hydration effects and the intra-peptide interactions leading to residue-residue contacts along the 
chain, although there are trapped water molecules. The interplay can be deduced from the  non-monotonous behavior of
the squared radius of gyration $R_g^2$  as function of time in Fig.\ \ref{snap_GL20}(b), obtained from $5$ independent runs. 
Note that for all the cases $R_g^2$ decays eventually to the equilibrium value. 

\par
In order to probe further the structural evolution of the chain along the collapse of $\textrm{(Gly)}_{20}$, we calculate the static structure factor $S(q)$ at different times.
Fig.\ \ref{snap_GL20}(c) shows $S(q)$ for the times corresponding to the snapshots. At $t=0$ ns, within the range $q \in [3,30]~\textrm{nm}^{-1}$, 
the chain can be described as an extended coil with $S(q) \sim q^{-1/\nu}$ \cite{rubenstein2003}, where $\nu=3/5$ is the critical 
exponent describing the scaling of $R_g \sim N^{\nu}$ for a self-avoiding polymer. With time  the decay exponent should increase
 from $-5/3$ and is expected to approach $-4$, in order to be consistent with the globule-like behavior of 
$S(q) \sim q^{-4}$ \cite{rubenstein2003}. Although the slope in our data in Fig.\ \ref{snap_GL20}(c) gradually increases with time, it 
does not appear to approach $-4$. This again could be due to the still ongoing interplay between the hydration effect and the intra-peptide interactions which hinders   the chain to form a compact globule, however, extending the simulations  up to $20$ ns does not change the overall behavior. Similar observations are made for all systems $\textrm{(Gly)}_{N}$ having a chain length of $N<50$ residue units. 
\par
\begin{figure}
\centering
\includegraphics[width=0.69\textwidth]{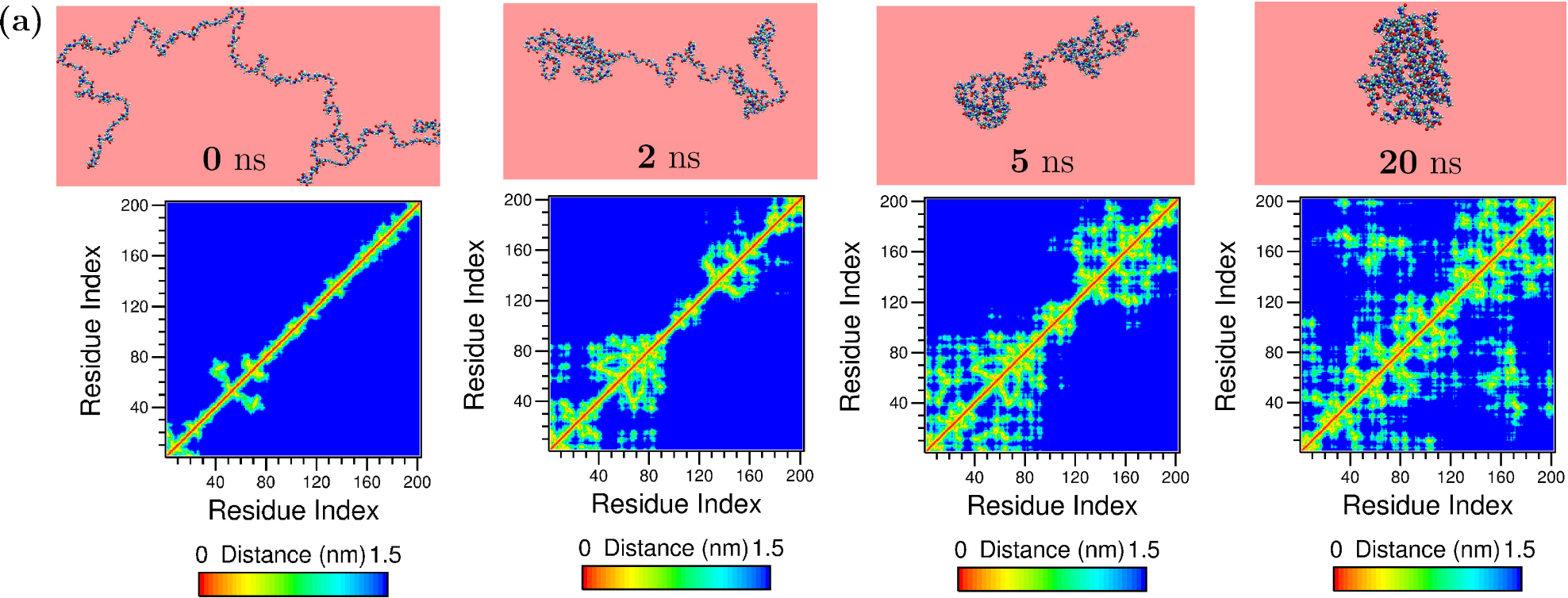}
\includegraphics[width=0.29\textwidth]{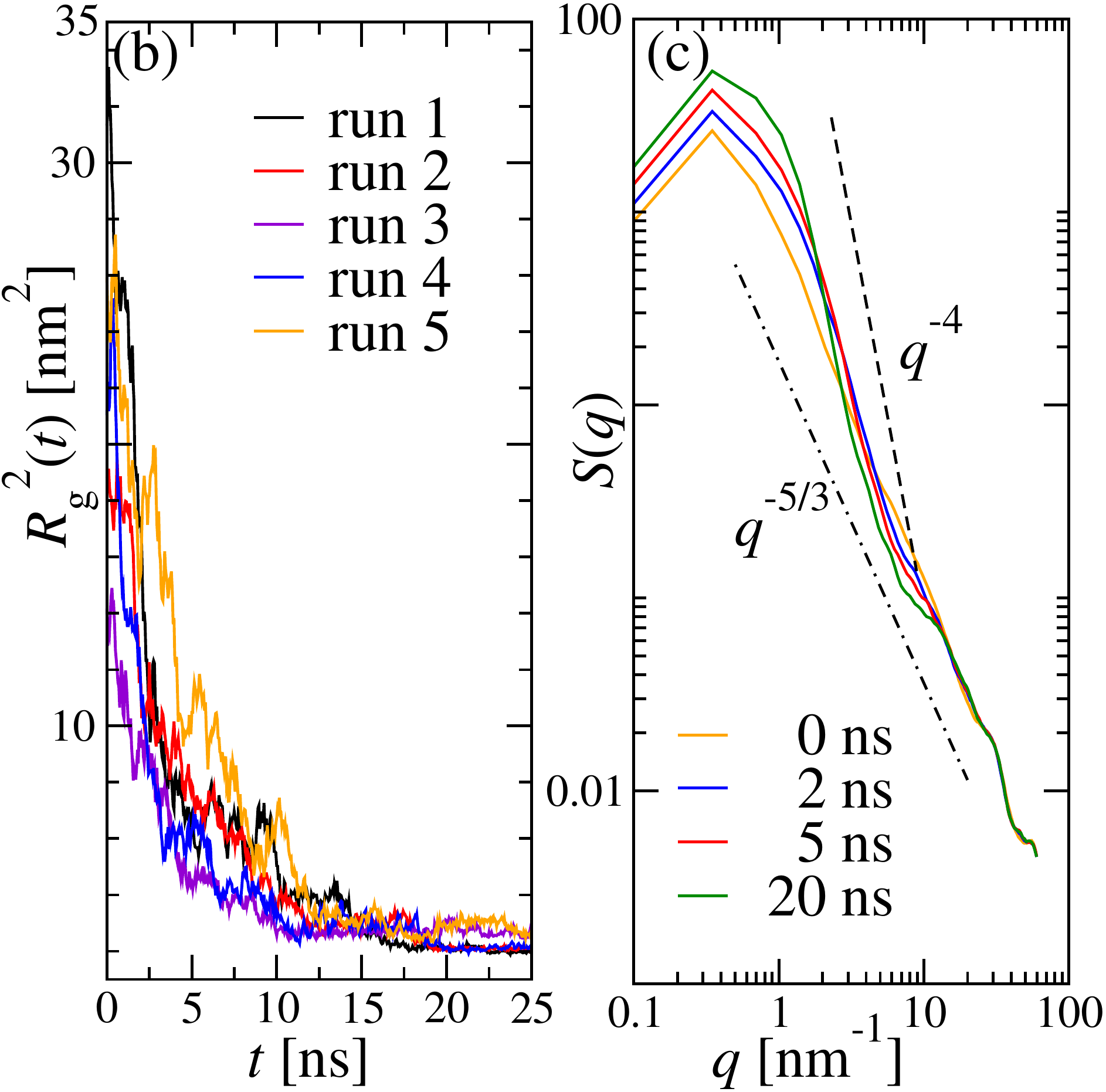}
\caption{\textbf{Pearl-necklace formation during collapse of longer chains.} (a) Same as in Fig.\ \ref{snap_GL20}(a) but for $\textrm{(Gly)}_{200}$ and correspondingly at different times, as mentioned. 
(b) Time dependence of the squared radius of gyration $R_g^2(t)$ from $5$ different runs. (c) The structure factors $S(q)$ at 
times that are presented in (a). There dashed lines have the same meaning as in  Fig.\ \ref{snap_GL20} (c). }
\label{snap_GL2h}
\end{figure}

For longer chains, the collapse is more pronounced, and  we finally  encounter  
characteristic features reminiscent of the homopolymer collapse. For instance,
in Fig.\ \ref{snap_GL2h}(a), we present  snapshots of the collapse of $\textrm{(Gly)}_{200}$ at $T_q=290$ K.  
The sequence of these snapshots  demonstrates a process that starts with local ordering of the residues along the chain. These local structures later merge with each other before finally forming a single globule at $t\approx 20$ ns. The emergence of these local arrangements 
is similar to   the  formation of local clusters  in the ``pearl-necklace'' picture of homopolymer collapse  \cite{Halperin2000,byrne2000,MajumderEPL,majumder2017SM}. The resemblance becomes even more obvious
 when looking at the corresponding contact maps  
in the lower panel. The box-like clustering along the diagonal indicates formation of ``pearls'' along the chain (see particularly at $t=2$ and $5$ ns)
that are reminiscent to the ones observed during the collapse of semiflexible homopolymer in Ref.\ \cite{lappala2013}. However, 
we do not see the anti-parallel hairpins that were associated with this diamond-shaped internal orders within these boxes.  
\par
In order to check for the presence of a competition between  hydration effects and the intra-peptide interactions  we probe again
 the time dependence of $R_g^2$ as measured in  five independent 
runs. Data are presented in Fig.\ \ref{snap_GL2h}(b). Unlike for the shorter  $\textrm{(Gly)}_{20}$ chain, 
the radius of gyration is now monotonically decreasing. This can be explained by the assumption 
 that for longer chains the intra-chain interactions  overcome the hydration effects. 
A similar picture emerges from Fig.\ \ref{snap_GL2h}(c).  The  plots of 
the structure factor $S(q)$ as function of time  demonstrate  how 
the extended coil behavior of $S(q) \sim q^{-5/3}$ at $t=0$ ns gradually changes to a globule-like behavior of $S(q) \sim q^{-4}$ at $t=20$ ns.

\begin{figure}
\centering
\includegraphics[width=0.85 \textwidth]{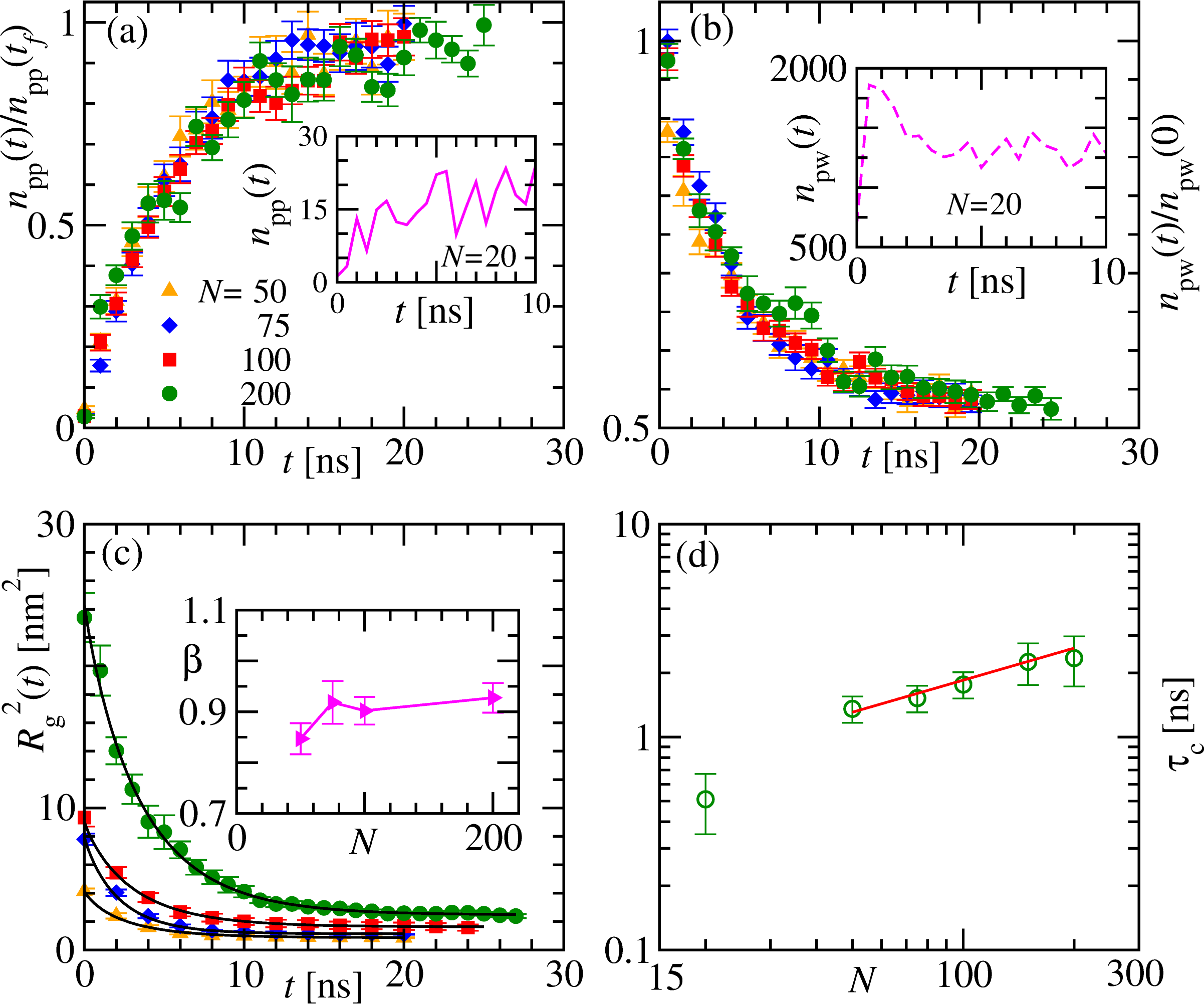}
\caption{\textbf{Kinetics of H-bonding and scaling of the collapse time.} (a) Time dependence of the number of protein-protein hydrogen bonds $n_{\textrm{pp}}(t)$ during collapse of $\textrm{(Gly)}_{N}$ for different $N$. To make the curves
fall within the same scale the data is normalized with $n_{\textrm{pp}}(t_f)$; $t_f$ is the maximum run time the simulations are done. The inset shows
time dependence of $n_{\textrm{pp}}(t)$ for $\textrm{(Gly)}_{20}$. (b) Same as in (a) but for the number of protein-water hydrogen bonds $n_{\textrm{pw}}(t)$. The inset again is 
same as the inset of (a) but for $n_{\textrm{pw}}(t)$. (c) Variation of the average squared radius of gyration $R_g^2(t)$ with time for the same systems that are 
presented in (a) and (b). The solid black lines are respective fits using the form  \eqref{rg_decay} and the corresponding $\beta$ obtained is shown as a function of $N$ 
in the inset. (d) Dependence of the collapse times $\tau_c$, extracted from the time decay of $R_g^2$ on the number of residue $N$. The solid line represents
the behavior $\tau_c \sim N^z$ with $z=0.5$.}
\label{hbond-relax}
\end{figure}
\par
Next, we analyze the number of intra-molecular, i.e., protein-protein $n_{\textrm{pp}}$ and 
inter-molecular, i.e., protein-water $n_{\textrm{pw}}$ hydrogen (H)-bonds. 
These quantities, measured for different $N$ and normalized by  the respective values at $t_f$
 (the maximum time up to which the simulations are run; for details see the 
Method section)  are plotted as function of time  in the main frame of Fig.\ \ref{hbond-relax}(a) and (b). 
Data for all $N$ in Fig.\ \ref{hbond-relax}(a) attain the value $\approx 1$ at the same time, demonstrating
 a reasonable overlap of the normalized  data. Similarly happens 
in (b) the decay of $n_{\textrm{pw}}$ to the saturation value at almost  the same time for different $N$, 
leading again  to  nicely overlapping curves. In the inset 
of Fig.\ \ref{hbond-relax}(a),  the time dependence of $n_{\textrm{pp}}$ for $\textrm{(Gly)}_{20}$ is non-monotonous  whereas 
the  $n_{\textrm{pw}}$ data  in the inset of Fig.\ \ref{hbond-relax}(b) exhibit a jump at early time before reaching saturation. This  again 
confirms the hydration effects for smaller chains. The overlap of the hydrogen-bond kinetics for large $N$ indicates that the collapse is not 
guided by the intra-peptide hydrogen bonds but depends mostly on the intra-peptide van der Waals interactions. 

However, the overlap of the hydrogen-bond data  does not allow one to calculate the collapse time $\tau_c$ from the time evolution of this quantity. 
More suitable for this purpose is the decay of the average squared radius of gyration $R_g^2$ depicted in Fig.\ \ref{hbond-relax}(c).
 The non-overlapping  data are  consistent with the respective solid lines 
obtained from the previously proposed fit \cite{majumder2017SM,christiansen2017JCP}
\begin{equation}\label{rg_decay}
 R_{g}^2(t)=b_0+b_1\exp[-(t/\tau_c)^{\beta}],
\end{equation}
where $b_0$ corresponds to the value of $R_{g}^2(t)$ in the collapsed state, and 
$b_1$ and $\beta$ are associated non-trivial fitting parameters. The obtained values of 
$\beta$ [see the inset of Fig.\ \ref{hbond-relax}(c)] indicate a very weak dependence on $N$, 
similar to the case of the earlier studied collapse of synthetic homopolymers \cite{majumder2017SM}. 
Although the above fit yields a collapse time $\tau_c$, more accurate estimates can be calculated 
  from the time when $R_{g}^2(t)$ has decayed to $50\%$  of its total decay, i.e., $\Delta R_{g}^2= R_{g}^2(0)-R_{g}^2(t_f)$. 
We plot the measured values of $\tau_c$ for different chain length  $N$ (including $N=20$) in Fig.\ \ref{hbond-relax}(d)
to check for a scaling of the form   $\tau_c \sim N^z$. Due to the competition between  hydration effects and intra-peptide
 interactions that dominate for smaller $N$ one expects  distinct scaling forms
 for small and large $N$. Our data indeed hint at the existence of two such scaling regions. Especially interesting is the consistency 
of our data for large $N$ with the solid line having $z=0.5$. This exponent  suggests that the dynamics is faster than 
the one  observed in MC simulations of non-biological homopolymer \cite{majumder2017SM}. Surprisingly, it is even faster than in the case of 
homopolymer collapse in presence of hydrodynamics \cite{Abrams2002,yeomans2005}. We conjecture that the more rapid 
collapse is due to the almost instantaneous presence of intra-chain hydrogen bonds that hasten local ordering.
 Simulations of longer chains would be desirable to confirm the value of $z=0.5$ and the super-fast collapse mechanism in
hydrogen-bonded polymers, however, such
simulations were computationally too costly to be considered in the present study. 

\par
\begin{figure}
\centering
\includegraphics[width=0.85 \textwidth]{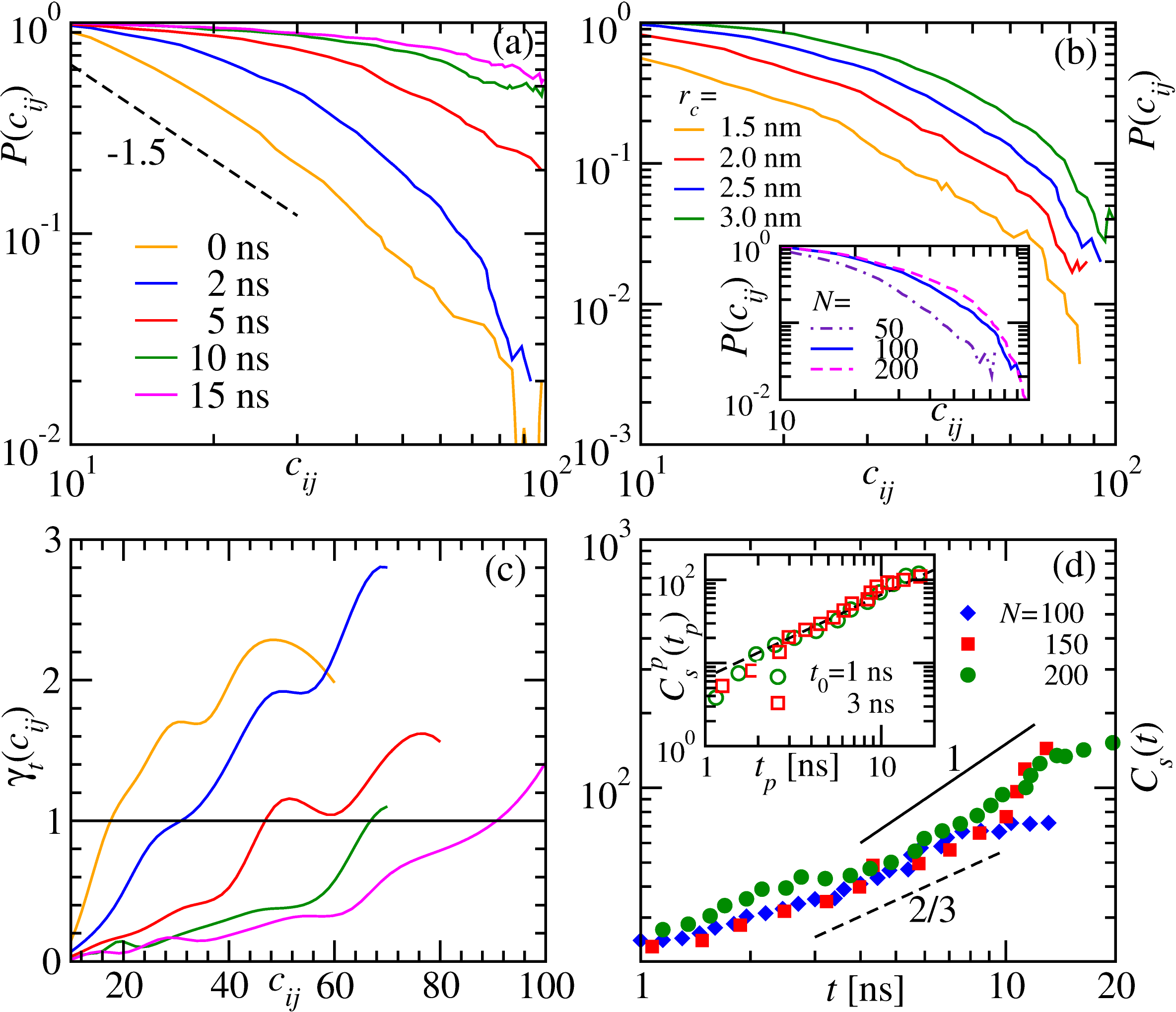}
\caption{\textbf{Cluster growth during the collapse.} (a) Contact probability $P(c_{ij})$ calculated using the cut-off $r_c=2.5$ nm, as 
a function of the distance $c_{ij}$ along the chain, at 
five different times during collapse of $\textrm{(Gly)}_{100}$. (b) Shows the consistency or the proportionality behavior of estimated 
contact probability $P(c_{ij})$ at a fixed time $t=2$ ns using different $r_c$ as indicated. The inset shows $P(c_{ij})$ at $t=2$ ns using $r_c=2.5$ nm 
for different $N$. (c) The discrete slope $\gamma_t$ obtained from Eq.\ \eqref{gammat} as a function of $c_{ij}$ for the times presented in (a). The solid 
line is for $\gamma_t=1$. (d) The main frame shows the growth of the mean cluster or pearl size $C_s(t)$ with time for different $N$. The solid lines 
represent power-law behavior $C_s(t) \sim t^{\alpha}$ with $\alpha=1$ and $2/3$, respectively. The inset shows the plot of $C_s^p(t_p)$ as function 
of the shifted time $t_p=t-t_0$ for two different choices of $t_0$. The solid line there represents a linear behavior.}
\label{pearl-kinetics}
\end{figure}

In a final step we want to quantify the coarsening kinetics of the ``pearl-necklace''  observed in Fig.\ \ref{snap_GL2h}(a). A measure 
of the relevant length scale, i.e., the mean cluster or pearl size $C_s(t)$, can be obtained from a box-plot analysis of the contact maps \cite{lappala2013}. 
Conjecturing that the collapse is driven by the intra-peptide van der Waals attraction of the backbone, we  extract
$C_s(t)$ from an  analysis of the contact probability $P(c_{ij})$ as a function of the contour distance $c_{ij}=|i-j|$ between any two 
C$\alpha$-atoms at the $i$-th and $j$-th position along the chain \cite{Scolari2018}. Two C$\alpha$-atoms are said to have a contact if they 
are within a cut-off distance $r_c$. Using $r_c=2.5$ nm, we show in Fig.\ \ref{pearl-kinetics}(a) values of   $P(c_{ij})$ calculated at different 
times during the collapse of  $\textrm{(Gly)}_{100}$. These contact probabilities  indicate  indeed
 a growing length scale as their decay slows with time. At the beginning, for $t=0$ ns, the chain is in extended state and $P(c_{ij})$ decays according 
to a power law  $P(c_{ij}) \sim c_{ij}^{-\gamma}$ with an exponent $\gamma=1.5$, as expected in a good solvent \cite{deGennesbook}. As time 
progresses, this power-law behavior appears at larger $c_{ij}$ after crossing over from a plateau-like behavior for small $c_{ij}$ which marks the 
local ordering along the chain. For  any reasonable choice of 
$r_c$ the form of the curves stays unchanged as demonstrated in Fig.\ \ref{pearl-kinetics}(b). Similarly, the form of the curve also does not depend
on the chain length  $N$  as illustrated in the inset of Fig.\ \ref{pearl-kinetics}(b) where we use $r_c=2.5$ nm and choose the point in 
 time $t=2$ ns. 

\par
The crossover point in the decay $P(c_{ij})$ as a function of $c_{ij}$ is  estimated from the discrete local slope as calculated by \cite{Scolari2018}
\begin{equation}\label{gammat}
 \gamma_t(c_{ij})= \frac{\Delta \ln [P(c_{ij})]}{\Delta \ln [c_{ij}]}.
\end{equation}
Plots of $\gamma_t(c_{ij})$ as a function of $c_{ij}$ are shown Fig.\ \ref{pearl-kinetics}(c) for the data presented in Fig.\ \ref{pearl-kinetics}(a). The 
crossing of the data with the $\gamma_t=1$ line happens at larger $c_{ij}$ as $t$ increases, and thus this crossover point gives a measure of $C_s(t)$. 
The obtained $C_s(t)$ for three different $N$ are shown as a function of $t$ on a 
double-log scale in the main frame of Fig.\ \ref{pearl-kinetics}(d). The flattening of the data at very large $t$ is due to finite-size effects when 
no more ordering is possible due to formation of single globule. At large $t$, before hitting size effects, the growth resembles a 
power law $C_s(t)=A_Nt^{\alpha}$ where the amplitude $A_N$ depends still on the chain length $N$ as the considered $N$ are not large enough.
 Hence,  $P(c_{ij})$ calculated using the same $r_c$ will overlap with each other, a fact that is demonstrated in the inset of Fig.\ \ref{pearl-kinetics}(b). However, since their form stays invariant in the large $t$ regime, they apparently 
follow the same power law. Our data do not span over decades in time, hence, it is hard to distinguish between $\alpha=2/3$ and $1$ behavior 
as shown by the dashed and the solid lines, respectively. In such cases it is advantageous  to describe the growth instead as
 $C_s(t)=C(t_0)+A_N(t-t_0)^{\alpha}$ by considering 
a crossover time $t_0$ and cluster size $C(t_0)$. This ansatz, originally developed for ferromagnets \cite{majumder_Ising},  was already 
 necessary in our earlier work for describing the  collapse of synthetic  homopolymers \cite{MajumderEPL,majumder2017SM,christiansen2017JCP}. 
Using the transformation $C_s^p(t_p)=C_s(t)-C(t_0)$ one finds $C_s^p(t_p)=At_p^\alpha$, with 
the shifted time $t_p=t-t_0$. For $\alpha=1$, the above transformation is invariant under any choice of $t_0$ in the post-crossover regime. This is demonstrated 
in the inset of Fig.\ \ref{pearl-kinetics}(d). The consistency of our data with the solid line representing a linear behavior further consolidates our finding of a 
linear cluster growth.
\par
In summary, we have investigated the nonequilibrium pathways by which polyglycine collapses in water. For  
short chains, the pathway  has few noticeable features and is driven by the competition between the hydration of the peptide, opposing the 
collapse, and the intra-peptide attractions, favoring the collapse \cite{asthagiri2017}. For long enough chains the importance of hydration
effects decreases, and   the kinetics 
of hydrogen bonds indicates that  van der Waals interactions of the backbone dominate  and drive the collapse. 
The nonequilibrium intermediates seen during the collapse exhibit local ordering or clustering 
that is analogous to the phenomenological ``pearl-necklace'' picture known to be valid 
for the earlier studied coarse-grained homopolymer models \cite{Halperin2000}. 
Using the contact probability of the C$\alpha$-atoms in the backbone, we extract a relevant dynamic length scale, i.e., cluster 
size, that as in simple homopolymer models grows  linearly with time \cite{majumder2017SM}. We believe that this linear growth is
  a result of the Brownian motion of the clusters and subsequent 
coalescence as in the case of droplet growth in fluids \cite{Binder1974}.
\par
Especially intriguing is that the scaling of the collapse time with length of the chain indicates  a faster dynamics, 
with a critical exponent $z = 0.5$ instead of
$z=1$, as seen in earlier homopolymer collapse studies \cite{Abrams2002,yeomans2005} that considered 
 simplified models describing non-hydrogen-bonded polymers such as polyethylene and polystyrene \cite{kremer1990}.
 The smaller exponent suggests that the 
fast folding times of proteins (typically in the $\mu$s -- ms range for proteins with $\approx 100-200$ residues) may be connected with 
a mechanism that allows in amino acid based polymers a more rapid  collapse than seen   in non-biological homopolymers, 
where  collapse times of $\approx 300$ ms \cite{Xu2006} up to $\approx 350$s \cite{chu1995} for  poly(N-isoporpylacrylamide) and polystyrene, respectively, have been reported.
This would also have implications for possible folding mechanisms as the fast collapse times in our simulations are connected with a quick appearance of local ordering. In fact, we conjecture that the smaller exponent $z$ is characteristic for collapse transitions where the presence
of intra-chain hydrogen bonding in amino acid based polymers immediately seeds (transient) local ordering, a step that in non-hydrogen-bonded
polymers only happens as the consequence of diffusive motion.  However, in order to test this conjecture, one would need to repeat first our above investigation for the other 19 amino acids. While such study is beyond the scope of
our current paper, the presented results demonstrate already that our approach provides a general platform to understand various conformational transitions  
that occur in biomolecules via local ordering. Another example would be, for instance, the helix-coil transition of polyalanine where 
the short-time dynamics has already been explored \cite{Arashiro2006,Arashiro2007}, or the study of two-time properties such as aging and dynamical scaling
in collapse and folding  \cite{Majumder2016PRE,christiansen2017JCP}.

\section*{Methods}
We construct $\textrm{(Gly)}_N$ molecules with hydrogenated N-terminus (--NH2) and C-terminus (--COOH).
 All-atom MD simulations are performed using standard 
GROMACS 5.0.2 tools while CHARMM22 with CMAP corrections \cite{mackerell1998,mackerell2004} is used for interactions between the atoms. For studying the 
collapse dynamics, we first prepare an extended chain in gas phase at $1500$ K. This follows solvation of this extended chain in a simple cubic box with water 
(modeled by the TIP3P model \cite{jorgensen1983}). The final MD run is performed at the desired quench temperature $T_q=290$ K which is lower than $310$ K, 
roughly the collapse transition temperature of $\textrm{(Gly)}_N$ in water. The size of the box and the number of water molecules, of course, are dependent on $N$ 
and are so chosen that the number density of water molecules is same for all $N$. For the smallest $N$, i.e., for $N=20$ the default box size was $4.2$ nm. 
Subsequently, the box sizes for longer chains were determined using the relation $R_g \sim N^{3/5}$. 
The size of the boxes should not have much role in collapse provided the two ends of the chain do not 
interact while using the periodic boundary condition. However, the number density of water molecules 
is supposed to play a role which we kept the same for all $N$. For $N=20$ the total number of water molecules 
used was $\approx 2000$ giving a number density of $\approx 32$ per $\textrm{nm}^{-3}$ which 
was maintained for all $N$. After the solvation we run our MD simulations using the Verlet-velocity integration 
scheme with time step $\delta t=2$ fs, in the NVT ensemble using 
the Nos\'{e}-Hoover thermostat that conserves the linear momentum, and thus is believed to be sufficient for 
preserving hydrodynamic effects \cite{frenkel_book}. 
Here, we use chains of length $N \in [20,50,75,100,150,200]$, and all the results presented are averaged over $50$ different initial configurations, 
except for $N=200$ where this number is $15$. The simulations are run up to time $t_f$ which is $10$ ns for $N=20$, $20$ ns for $N \in [50,150]$, and $25$ ns for $N=200$. 
\par

For a polymer of length $N$ (number of monomers) the squared radius of gyration is calculated as 
$R_g^2= \Sigma_{ij}(\vec{r}_i-\vec{r}_j)^2/2N^2$. For $\textrm{(Gly)}_{N}$ the chain length was determined from $N$, the number 
of residues or repeating units which contain a fixed set of atoms. Thus $R_g$ for $\textrm{(Gly)}_{N}$ was calculated 
considering all the atoms present in all the residues. However, the scaling can still be checked in terms $N$, as is 
done here. The structure factor for a polymer of length $N$ is calculated using the relation 
$S(\vec{q})=(1/N)\Sigma_{ij}\exp(-i\vec{q} \cdot \vec{r}_{ij})$, where $\vec{r}_{ij}$ is the distance vector between the $i$-th and $j$-th monomer along the chain. 
As explained above in the case for measuring $R_g$, for $S(q)$ too we use all the atoms in all the residues.
We calculate the hydrogen bonds using the standard GROMACS tool gmx hbond. It considers all possible donars 
and acceptors and decides for the existence of a hydrogen bond if the distance between them is less than $0.35$ nm and the 
hydrogen-donar-acceptor angle is less than $30$\textdegree.
%




\begin{acknowledgments}
This project was funded by the Deutsche Forschungsgemeinschaft (DFG) under Grant Nos. SFB/TRR 102 (project B04) and JA 483/33-1 and 
the National Institutes of Health (NIH) under grants GM120578 and GM120634.
 It was further supported by the Deutsch-Franz\"{o}sische Hochschule (DFH-UFA) 
through the Doctoral College ``$\mathbb{L}^4$'' under Grant No. CDFA-02-07 and the Leipzig Graduate School of Natural Sciences ``BuildMoNa''. 
U.H. thanks the Institut f\"{u}r Theoretische Physik and especially the Janke group for kind hospitality during his sabbatical stay at Universit\"{a}t Leipzig.
\end{acknowledgments}

\section*{Contributions}
All authors contributed to develop the project and S.M.
performed the simulations. All authors discussed, analyzed the results
and wrote the manuscript.
\section*{COMPETING FINANCIAL INTERESTS}
The authors declare that they have no
competing financial interests.

%

\end{document}